\documentclass[11pt,a4paper]{article}

\usepackage[a4paper,margin=1in]{geometry}
\usepackage{lmodern}
\usepackage{amsmath,amssymb,mathtools}
\usepackage{graphicx}
\usepackage{longtable,booktabs,array,calc}
\usepackage{caption}
\usepackage{fvextra}
\usepackage{enumitem}
\usepackage{microtype}
\usepackage{xurl}
\usepackage[hidelinks]{hyperref}
\usepackage{textcomp}
\allowdisplaybreaks[2]
\setlist{nosep}
\hypersetup{pdftitle={Bayesian approach to the Student's t-test based on the B-FBST},
            pdfauthor={Ahmed Hamimes}}

\newcommand{\real}[1]{#1}

\title{Bayesian approach to the Student\textquotesingle s t-test based on the B-FBST}
\author{Ahmed Hamimes\\[0.4em]
\small BIOSTIM Laboratory, Medicine Faculty Salah Boubnider Constantine 03 University Constantine, Algeria\\
\small \texttt{ahmed.hamimes@univ-constantine3.dz}}
\date{}

\begin{document}
\maketitle
\renewcommand{\abstractname}{Summary}
\begin{abstract}
This article proposes an exact Bayesian formulation of the B-FBST Student\textquotesingle s t-test applied to the one-sample Student\textquotesingle s t-test. The method relies on constructing an exact posterior distribution of the standardized parameter, then calculating the e-value to measure the evidence against the null hypothesis. To improve the stability of the procedure, posterior densities obtained from bootstrap samples are aggregated using random weights.

A Monte Carlo simulation study was then conducted to compare the B-FBST Student\textquotesingle s t-test with the classical Student\textquotesingle s t-test and the bootstrap Student\textquotesingle s t-test. The results, obtained from 100 Monte Carlo replications and 200 bootstrap replications, show that the B-FBST Student\textquotesingle s t-test exhibits favorable overall performance, with an average of 71.0\% correct decisions, compared to 65.5\% for the classical test and 70.7\% for the bootstrap test. The proposed method thus appears to be a robust alternative, particularly in several situations where the classical assumptions of the Student\textquotesingle s t-test are compromised.

\end{abstract}

\noindent\textbf{Keywords:} Student\textquotesingle s t-test, B-FBST, Bayesian approach, e-value, bootstrap, Monte Carlo simulation; hypothesis testing; posterior density.

\section{Introduction}

The study of a mean from a sample is a central problem in statistical inference. When the population standard deviation is known, comparison with a reference value can be based on a standard normal statistic (Z-test). However, in most real-world applications, this information is unavailable: the variance must then be estimated from the sample itself. This estimation introduces a source of uncertainty, leading to the framework of the Student\textquotesingle s t-test, introduced in the seminal article on the probable error of a mean (Student, 1908).

The Student\textquotesingle s t-test remains a standard method for analyzing means, but its use requires certain conditions, including the independence of observations and an approximately normal distribution, especially when the sample size is small. In real-world data, these conditions can be disrupted by extreme values, skewness, thick tails, heteroscedasticity, or a more complex data structure. These situations can reduce the stability of the classical test and justify the investigation of more robust alternative or complementary methods.

From this perspective, the Bayesian approach offers an interesting framework, as it allows for the description of statistical uncertainty based on a posterior distribution. This work adopts an objective Bayesian formulation, based on a non-informative prior distribution of Jeffreys. This choice aims to limit the influence of external information and to make the inference primarily dependent on the observed data (Jeffreys, 1961). The Full Bayesian Significance Test, or FBST, constitutes the main methodological framework of this study. This approach was proposed to test specific hypotheses without directly resorting to the frequentist logic of the p-value. The FBST is based on the e-value, which measures the statistical evidence associated with the hypothesis being tested based on the posterior density (Pereira \& Stern, 1999). More recent work also presents e-value as a Bayesian measure suitable for specific assumptions and highlights its properties of logical consistency and invariance (Pereira \& Stern, 2022; Stern et al., 2025).

The core of the proposed method relies on a bootstrap-Bayesian aggregation of exact posterior densities. From the initial sample, several bootstrap samples are generated. For each of these, the standardized parameter associated with the Student\textquotesingle s t-test is calculated, and then its exact posterior density is constructed. These densities are not used separately; they are combined into an aggregated posterior density using Dirichlet-type random weights. This combination allows for the integration of uncertainty related to resampling while preserving the exact Bayesian structure of the model. Thus, the e-value of the FBST is calculated not from a single posterior density, but from a more stable aggregate density that summarizes the information from the different bootstrap replications. This approach is inspired both by the bootstrap principle, used to evaluate the variability of estimators (Efron, 1979), and by the Bayesian bootstrap based on the use of random weights (Rubin, 1981).

The aim of this study is therefore to propose an exact Bayesian formulation of the B-FBST test applied to the Student\textquotesingle s t-test on a single sample, and then to evaluate its performance through simulation. The B-FBST Student\textquotesingle s t-test is compared to two reference methods: the classical Student\textquotesingle s t-test and the bootstrap Student\textquotesingle s t-test. The Monte Carlo simulation is based on 100 and 200 bootstrap replications, according to ten scenarios: normal case, Student\textquotesingle s t-distribution, contaminated data, centered log-normal distribution, centered nonlinear structure, large sample with small effect, small sample, heteroscedasticity, mixing of groups, and negative effect. This comparison allows us to analyze under what conditions the B-FBST Student\textquotesingle s t-test can constitute a robust alternative to the classical Student\textquotesingle s t-test.

\section{Student\textquotesingle s t-test}

Student\textquotesingle s t-test is a statistical method used to test a hypothesis about a mean or a difference of means. It is particularly useful when the population variance is unknown and must be estimated from the sample data.

\subsection*{Student\textquotesingle s t-test for a single sample}

We consider a sample:

\begin{equation}\label{eq:auto-001}
X_{1},X_{2},\ldots,X_{n}.
\end{equation}

We want to test whether the population mean is equal to a reference value \(\mu_{0}\). The hypotheses are:

\begin{equation}\label{eq:auto-002}
H_{0}:\mu = \mu_{0}
\end{equation}

against :

\begin{equation}\label{eq:auto-003}
H_{1}:\mu \neq \mu_{0}.
\end{equation}

The empirical mean is:

\begin{equation}\label{eq:auto-004}
\bar{X} = \frac{1}{n}\sum_{i = 1}^{n}X_{i}.
\end{equation}

The empirical standard deviation is:

\begin{equation}\label{eq:auto-005}
s = \sqrt{\frac{1}{n - 1}\sum_{i = 1}^{n}(X_{i} - \bar{X})^{2}}.
\end{equation}

The Student\textquotesingle s t-test statistic is then:

\begin{equation}\label{eq:auto-006}
t_{obs} = \frac{\bar{X} - \mu_{0}}{s/\sqrt{n}}.
\end{equation}

Under the null hypothesis \(H_{0}\), this statistic follows a Student\textquotesingle s t-distribution with \(n - 1\)n degrees of freedom:

\begin{equation}\label{eq:auto-007}
t_{obs} \sim t_{n - 1}.
\end{equation}

For a two-sided threshold test \(\alpha\), we reject \(H_{0}\)if:

\begin{equation}\label{eq:auto-008}
\lvert t_{obs} \rvert > t_{1 - \frac{\alpha}{2},\,n - 1}.
\end{equation}

\subsection*{Student\textquotesingle s t-test for two independent samples}

Student\textquotesingle s t-test can also be used to compare the means of two independent populations. We consider two samples:

\begin{equation}\label{eq:auto-009}
X_{1},\ldots,X_{n_{1}}
\end{equation}

And :

\begin{equation}\label{eq:auto-010}
Y_{1},\ldots,Y_{n_{2}}.
\end{equation}

Note:

\begin{equation}\label{eq:auto-011}
\bar{X},\bar{Y}
\end{equation}

the empirical means of the two groups, and:

\begin{equation}\label{eq:auto-012}
s_{X}^{2},s_{Y}^{2}
\end{equation}

their empirical variances.

The goal is to test:

\begin{equation}\label{eq:auto-013}
H_{0}:\mu_{1} = \mu_{2}
\end{equation}

against :

\begin{equation}\label{eq:auto-014}
H_{1}:\mu_{1} \neq \mu_{2}.
\end{equation}

When the variances of the two populations are assumed to be equal, the combined variance is used:

\begin{equation}\label{eq:auto-015}
s_{p}^{2} = \frac{(n_{1} - 1)s_{X}^{2} + (n_{2} - 1)s_{Y}^{2}}{n_{1} + n_{2} - 2}.
\end{equation}

The test statistic is then:

\begin{equation}\label{eq:auto-016}
t_{obs} = \frac{\bar{X} - \bar{Y}}{s_{p}\sqrt{\frac{1}{n_{1}} + \frac{1}{n_{2}}}}.
\end{equation}

Under the null hypothesis \(H_{0}:\mu_{1} = \mu_{2}\), this statistic follows a Student\textquotesingle s t-distribution:

\begin{equation}\label{eq:auto-017}
n_{1} + n_{2} - 2
\end{equation}

degrees of freedom:

\begin{equation}\label{eq:auto-018}
t_{obs} \sim t_{n_{1} + n_{2} - 2}.
\end{equation}

The decision rule for a two-sided threshold test \(\alpha\)is:

\begin{equation}\label{eq:auto-019}
\mid t_{obs} \mid > t_{1 - \alpha/2,n_{1} + n_{2} - 2} \Longrightarrow \text{rejet~de~}H_{0}.
\end{equation}

When the variances are not assumed to be equal, Welch\textquotesingle s test is used instead. In this case, the statistic is:

\begin{equation}\label{eq:auto-020}
t_{obs} = \frac{\bar{X} - \bar{Y}}{\sqrt{\frac{s_{X}^{2}}{n_{1}} + \frac{s_{Y}^{2}}{n_{2}}}}.
\end{equation}

The degrees of freedom are then approximated by Welch\textquotesingle s formula:

\begin{equation}\label{eq:auto-021}
\nu = \frac{\left( \frac{s_{X}^{2}}{n_{1}}+\frac{s_{Y}^{2}}{n_{2}} \right)^{2}}{\frac{\left( \frac{s_{X}^{2}}{n_{1}} \right)^{2}}{n_{1} - 1} + \frac{\left( \frac{s_{Y}^{2}}{n_{2}} \right)^{2}}{n_{2} - 1}}.
\end{equation}

We then have approximately:

\begin{equation}\label{eq:auto-022}
t_{obs} \sim t_{\nu}.
\end{equation}

\subsection*{Student\textquotesingle s t-test for paired observations}

When observations are matched, the two measurements are taken on the same individuals or on related units. This is the case, for example, with a measurement before and after a treatment.

We consider the following pairs of observations:

\begin{equation}\label{eq:auto-023}
(X_{i},Y_{i}),i = 1,\ldots,n.
\end{equation}

We define the differences:

\begin{equation}\label{eq:auto-024}
D_{i} = X_{i} - Y_{i}.
\end{equation}

The paired test therefore amounts to applying a Student\textquotesingle s t-test to a single sample on the differences \(D_{i}\).

The assumptions are:

\begin{equation}\label{eq:auto-025}
H_{0}:\mu_{D} = 0
\end{equation}

against :

\begin{equation}\label{eq:auto-026}
H_{1}:\mu_{D} \neq 0,
\end{equation}

Or \(\mu_{D}\)denotes the average of the differences in the population.

The average of the differences is:

\begin{equation}\label{eq:auto-027}
\bar{D} = \frac{1}{n}\sum_{i = 1}^{n}D_{i}.
\end{equation}

The empirical standard deviation of the differences is:

\begin{equation}\label{eq:auto-028}
s_{D} = \sqrt{\frac{1}{n - 1}\sum_{i = 1}^{n}(D_{i} - \bar{D})^{2}}.
\end{equation}

The paired Student\textquotesingle s t-test statistic is:

\begin{equation}\label{eq:auto-029}
t_{obs} = \frac{\bar{D}}{s_{D}/\sqrt{n}}.
\end{equation}

Under the null hypothesis \(H_{0}:\mu_{D} = 0\), this statistic follows a Student\textquotesingle s t-distribution with \(n - 1\)n degrees of freedom:

\begin{equation}\label{eq:auto-030}
t_{obs} \sim t_{n - 1}.
\end{equation}

The decision rule is:

\begin{equation}\label{eq:auto-031}
\mid t_{obs} \mid > t_{1 - \alpha/2,n - 1} \Longrightarrow \text{rejet~de~}H_{0}.
\end{equation}Student\textquotesingle s t-test relies on several assumptions. Observations must be independent in the case of a one- or two-sample independent test. For a paired test, it is the differences \(D_{i}\)that must be independent. The variable being studied must be quantitative, and the distribution must be approximately normal, especially when the sample size is small.

In the case of two independent samples, if the variances are assumed to be equal, the classic Student\textquotesingle s t-test with combined variance is used. If this assumption is doubtful, Welch\textquotesingle s test is preferable.

Thus, the Student\textquotesingle s t-test is a fundamental procedure for comparing a mean to a theoretical value, comparing two independent means, or analyzing paired data.

\section{Bayesian formulation of the single-sample Student\textquotesingle s B-FBST test}

We consider a random sample:

\begin{equation}\label{eq:auto-032}
Y_{1},\ldots,Y_{n} \mid \mu,\sigma^{2}\sim \mathcal{N}(\mu,\sigma^{2}),
\end{equation}

Or \(\mu\)is the unknown mean and \(\sigma^{2}\)the unknown variance. The objective is to test the point hypothesis:

\begin{equation}\label{eq:auto-033}
H_{0}:\mu = \mu_{0}
\end{equation}

against the bilateral hypothesis:

\begin{equation}\label{eq:auto-034}
H_{1}:\mu \neq \mu_{0}.
\end{equation}

Note:

\begin{equation}\label{eq:auto-035}
\bar{Y} = \frac{1}{n}\sum_{i = 1}^{n}Y_{i}
\end{equation}

And :

\begin{equation}\label{eq:auto-036}
S = \sum_{i = 1}^{n}(Y_{i} - \bar{Y})^{2}.
\end{equation}

The empirical variance is given by:

\begin{equation}\label{eq:auto-037}
s^{2} = \frac{S}{n - 1}.
\end{equation}

Within the Bayesian framework, we use Jeffreys\textquotesingle{} non-informative prior law:

\begin{equation}\label{eq:auto-038}
\pi(\mu,\sigma^{2}) \propto \frac{1}{\sigma^{2}}.
\end{equation}

After observing the data \(y = (y_{1},\ldots,y_{n})\), the conditional distribution \(\mu\)given \(\sigma^{2}\)is:

\begin{equation}\label{eq:auto-039}
\mu \mid \sigma^{2},y \sim \mathcal{N}\left( \bar{y},\frac{\sigma^{2}}{n} \right).
\end{equation}

Furthermore, the posterior marginal distribution of \(\sigma^{2}\)is an inverse-gamma distribution:

\begin{equation}\label{eq:auto-040}
\sigma^{2} \mid y \sim \text{Inverse-Gamma}\left( \frac{n - 1}{2},\frac{S}{2} \right).
\end{equation}

Its density is written as:

\begin{equation}\label{eq:auto-041}
\pi(\sigma^{2} \mid y) = \frac{\left( \frac{S}{2} \right)^{\frac{n - 1}{2}}}{\Gamma\left( \frac{n - 1}{2} \right)}(\sigma^{2})^{- \frac{n + 1}{2}}\exp\left( - \frac{S}{2\sigma^{2}} \right),\sigma^{2} > 0.
\end{equation}

By integrating \(\sigma^{2}\), we obtain the exact marginal distribution of \(\mu \mid y\). This distribution is a Student\textquotesingle s t-distribution:

\begin{equation}\label{eq:auto-042}
\mu \mid y \sim t_{n - 1}\left( \bar{y},\frac{s}{\sqrt{n}} \right).
\end{equation}

Its density is:

\begin{equation}\label{eq:auto-043}
\ \ \ \ \ \ \ \ \ \ \ \ \ \ \ \ \ \ \ \pi\left( \mu\mid y \right) = \frac{\Gamma\left( \frac{n}{2} \right)}{\Gamma\left( \frac{n - 1}{2} \right)\sqrt{(n - 1)\pi}\left( \frac{s}{\sqrt{n}} \right)}\left\lbrack 1+\frac{1}{n - 1}\left( \frac{\mu - \bar{y}}{\frac{s}{\sqrt{n}}} \right)^{2} \right\rbrack^{- \frac{n}{2}}.
\end{equation}

To apply FBST, we introduce the standardized parameter:

\begin{equation}\label{eq:auto-044}
\delta = \frac{\mu - \mu_{0}}{s}.
\end{equation}

Under the null hypothesis \(H_{0}:\mu = \mu_{0}\), we then have:

\begin{equation}\label{eq:auto-045}
H_{0}:\delta = 0.
\end{equation}

This transformation offers three advantages:

\begin{enumerate}
\def\labelenumi{\arabic{enumi}.}
\item
  It allows the difference between the unknown mean and the reference value to be expressed in a standardized unit.
\item
  It makes the parameter dimensionless, regardless of the unit of measurement of the observations.
\item
  It allows the point hypothesis to be formulated in a particularly simple form for the FBST: \(\delta = 0\).
\item
  It gives good simulation results compared to classical statistics.
\end{enumerate}

The empirical value associated with this parameter is:

\begin{equation}\label{eq:auto-046}
\widehat{\delta} = \frac{\bar{y} - \mu_{0}}{s}.
\end{equation}

Following \(\mu \mid y\)Student\textquotesingle s t-distribution, we directly obtain:

\begin{equation}\label{eq:auto-047}
\delta \mid y \sim t_{n - 1}\left( \widehat{\delta},\frac{1}{\sqrt{n}} \right).
\end{equation}

The exact density of \(\delta \mid y\)is therefore:

\begin{equation}\label{eq:auto-048}
\ \ \ \ \ \ \ \ \ \ \ \ \ \ \ \ \ \ \ \ \ \ \ \ \ \ \ \ \ \ \ \ \ \pi\left( \delta\mid y \right) = \frac{\Gamma\left( \frac{n}{2} \right)}{\Gamma\left( \frac{n - 1}{2} \right)\sqrt{(n - 1)\pi}\left( \frac{1}{\sqrt{n}} \right)}\left\lbrack 1+\frac{1}{n - 1}\left( \frac{\delta - \widehat{\delta}}{\frac{1}{\sqrt{n}}} \right)^{2} \right\rbrack^{- \frac{n}{2}}
\end{equation}

To account for sampling instability, \(R\)bootstrap samples are generated:

\begin{equation}\label{eq:auto-049}
y^{*(1)},\ldots,y^{*(R)}.
\end{equation}

For each bootstrap sample \(b = 1,\ldots,R\), we calculate:

\begin{equation}\label{eq:auto-050}
{\widehat{\delta}}^{*(b)} = \frac{{\bar{y}}^{*(b)} - \mu_{0}}{s^{*(b)}}.
\end{equation}

The corresponding bootstrapped density is then:

\begin{equation}\label{eq:auto-051}
\ \ \ \ \pi^{(b)}\left( \delta\mid y^{*(b)} \right) = \frac{\Gamma\left( \frac{n}{2} \right)}{\Gamma\left( \frac{n - 1}{2} \right)\sqrt{(n - 1)\pi}\left( \frac{1}{\sqrt{n}} \right)}\left\lbrack 1+\frac{1}{n - 1}\left( \frac{\delta - {\widehat{\delta}}^{*(b)}}{\frac{1}{\sqrt{n}}} \right)^{2} \right\rbrack^{- \frac{n}{2}}
\end{equation}

Next, random weights are introduced:

\begin{equation}\label{eq:auto-052}
(w_{1},\ldots,w_{R}) \sim \text{Dirichlet}(1,\ldots,1),
\end{equation}

with :

\begin{equation}\label{eq:auto-053}
w_{b} \geq 0,\sum_{b = 1}^{R}w_{b} = 1.
\end{equation}

The aggregated posterior density used in the B-FBST method is defined by:

\begin{equation}\label{eq:auto-054}
\pi^{(B)}(\delta \mid y) = \sum_{b = 1}^{R}w_{b}\pi^{(b)}(\delta \mid y^{*(b)}).
\end{equation}

This aggregate density is then used to apply the FBST. Under the null hypothesis, the density is evaluated at the point:

\begin{equation}\label{eq:auto-055}
\delta = 0.
\end{equation}

We therefore define:

\begin{equation}\label{eq:auto-056}
\pi_{0} = \pi^{(B)}(0 \mid y).
\end{equation}

The tangential region is:

\begin{equation}\label{eq:auto-057}
T = \left\{ \delta:\pi^{(B)}(\delta \mid y) > \pi^{(B)}(0 \mid y) \right\}.
\end{equation}

The evidence against the null hypothesis is given by:

\begin{equation}\label{eq:auto-058}
\bar{ev}(H_{0}) = \int_{T}^{}\pi^{(B)}(\delta \mid y)\,d\delta.
\end{equation}

The e-value associated with \(H_{0}\)is then:

\begin{equation}\label{eq:auto-059}
EV(H_{0}) = 1 - \bar{ev}(H_{0}).
\end{equation}

The decision rule adopted is:

\begin{equation}\label{eq:auto-060}
EV(H_{0}) < 0.20 \Longrightarrow \text{rejet~de~}H_{0}.
\end{equation}

Thus, a low value of \(EV(H_{0})\)indicates that the aggregated posterior density assigns more plausibility to values of that \(\delta\)are more consistent with the data than the value of zero \(\delta = 0\). This reflects evidence against the hypothesis \(H_{0}:\mu = \mu_{0}\).

\section{Simulation plan for the one-sample Student\textquotesingle s t-B-FBST test}

To evaluate the performance of the Bayesian B-FBST test applied to the Student\textquotesingle s t-test on a single sample, a Monte Carlo simulation study was conducted. The objective was to compare the proposed method to two reference methods: the classical Student\textquotesingle s t-test and the bootstrap Student\textquotesingle s t-test.

The test focuses on the null hypothesis:

\begin{equation}\label{eq:auto-061}
H_{0}:\delta = 0
\end{equation}

against the alternative bilateral hypothesis:

\begin{equation}\label{eq:auto-062}
H_{1}:\delta \neq 0.
\end{equation}

In this study, the standardized parameter is defined by:

\begin{equation}\label{eq:auto-063}
\delta = \frac{\mu - \mu_{0}}{s},
\end{equation}

Or \(\mu\)represents the population mean, \(\mu_{0}\)the tested reference value, and \(s\)the empirical standard deviation. In the simulations, the reference value is set to:

\begin{equation}\label{eq:auto-064}
\mu_{0} = 0.
\end{equation}

For each bootstrap sample \(b = 1,\ldots,R\), we calculate:

\begin{equation}\label{eq:auto-065}
{\widehat{\delta}}^{*(b)} = \frac{{\bar{y}}^{*(b)} - \mu_{0}}{s^{*(b)}}.
\end{equation}

The posterior distribution used for the standardized parameter is then given by:

\begin{equation}\label{eq:auto-066}
\delta \mid y^{*(b)} \sim t_{n - 1}\left( {\widehat{\delta}}^{*(b)},\frac{1}{\sqrt{n}} \right).
\end{equation}

Thus, each bootstrap sample provides a posteriori density denoted:

\begin{equation}\label{eq:auto-067}
\pi^{(b)}(\delta \mid y^{*(b)}).
\end{equation}

The different bootstrapped densities are then combined using random weights:

\begin{equation}\label{eq:auto-068}
(w_{1},\ldots,w_{R}) \sim \text{Dirichlet}(1,\ldots,1),
\end{equation}

with :

\begin{equation}\label{eq:auto-069}
w_{b} \geq 0,\sum_{b = 1}^{R}w_{b} = 1.
\end{equation}

The aggregated posterior density of the B-FBST method is defined by:

\begin{equation}\label{eq:auto-070}
\pi^{(B)}\left( \delta\mid y \right) = \sum_{b = 1}^{R}w_{b}\pi^{(b)}\left( \delta\mid y^{*(b)} \right).
\end{equation}

This aggregate density is then used to apply the FBST. Under the null hypothesis, the density is evaluated at the point \(\delta = 0\), and then the tangential region is defined:

\begin{equation}\label{eq:auto-071}
T = \left\{ \delta:\pi^{(B)}(\delta \mid y) > \pi^{(B)}(0 \mid y) \right\}.
\end{equation}

The e-value associated with the null hypothesis is given by:

\begin{equation}\label{eq:auto-072}
EV(H_{0}) = 1 - \int_{T}^{}\pi^{(B)}(\delta \mid y)\,d\delta.
\end{equation}

The decision rule adopted is:

\begin{equation}\label{eq:auto-073}
EV(H_{0}) < 0.20 \Longrightarrow \text{rejet~de~}H_{0}.
\end{equation}

Otherwise, the null hypothesis is not rejected.

The simulation is conducted with:

\begin{equation}\label{eq:auto-074}
MC = 100
\end{equation}

replications and:

\begin{equation}\label{eq:auto-075}
R = 200
\end{equation}

replications . For each scenario, the three methods are applied and the average percentage of correct decisions is calculated. A decision is considered correct when the method does not reject \(H_{0}\)when \(\mu = \mu_{0}\), and when it rejects \(H_{0}\)when \(\mu \neq \mu_{0}\).

Ten scenarios are considered in order to evaluate the methods in various statistical situations. The scenario \(S1\)corresponds to the normal reference case, where the data are generated according to:

\begin{equation}\label{eq:auto-076}
Y_{i}\sim \mathcal{N}(\mu,1),n = 30,\mu = 0.50.
\end{equation}

The scenario \(S2\)introduces a thick-tailed distribution:

\begin{equation}\label{eq:auto-077}
Y_{i} = \mu + T_{i},T_{i} \sim t_{3},n = 30,\mu = 0.50.
\end{equation}

The scenario \(S3\)considers contamination by atypical values:

\begin{equation}\label{eq:auto-078}
Y_{i} \sim (1 - \varepsilon)\mathcal{N}(\mu,1) + \varepsilon\mathcal{N}(\mu + 5,9),
\end{equation}

with :

\begin{equation}\label{eq:auto-079}
\varepsilon = 0.10,n = 30,\mu = 0.50.
\end{equation}

The scenario \(S4\)introduces a centered log-normal distribution:

\begin{equation}\label{eq:auto-080}
{Z_{i}\sim \mathcal{N}(0,1),
}{Y_{i} = \exp(Z_{i}) - \exp\left( \frac{1}{2} \right) + \mu,n = 50,\mu = 0.50.}
\end{equation}

The scenario \(S5\)corresponds to a centered non-linear structure:

\begin{equation}\label{eq:auto-081}
{X_{i}\sim \mathcal{N}(0,1),
}{Y_{i} = X_{i}^{2} - 1 + \varepsilon_{i},\varepsilon_{i}\sim \mathcal{N}(0,{0.5}^{2}),}
\end{equation}

with :

\begin{equation}\label{eq:auto-082}
n = 50,\mu = 0.
\end{equation}

In this case, the expected value of \(Y_{i}\)is zero; the correct expected decision is therefore the non-rejection of \(H_{0}\).

The scenario \(S6\)studies a large sample with a small effect:

\begin{equation}\label{eq:auto-083}
Y_{i}\sim \mathcal{N}(0.10,1),n = 200.
\end{equation}

The scenario \(S7\)corresponds to a small sample:

\begin{equation}\label{eq:auto-084}
Y_{i}\sim \mathcal{N}(0.50,1),n = 10.
\end{equation}

The scenario \(S8\)introduces heteroscedasticity:

\begin{equation}\label{eq:auto-085}
{X_{i}\sim \mathcal{N}(0,1),
}{Y_{i} = \mu + \sqrt{1 + X_{i}^{2}}\,\varepsilon_{i},\varepsilon_{i}\sim \mathcal{N}(0,1),}
\end{equation}

with :

\begin{equation}\label{eq:auto-086}
n = 50,\mu = 0.50.
\end{equation}

In this case :

\begin{equation}\label{eq:auto-087}
Var(Y_{i} \mid X_{i}) = 1 + X_{i}^{2}.
\end{equation}

The scenario \(S9\)involves a mix of groups. We generate:

\begin{equation}\label{eq:auto-088}
G_{i} \sim \text{Bernoulli}(0.30).
\end{equation}

If so \(G_{i} = 0\), then:

\begin{equation}\label{eq:auto-089}
Y_{i}\sim \mathcal{N}(\mu,1).
\end{equation}

If so \(G_{i} = 1\), then:

\begin{equation}\label{eq:auto-090}
Y_{i}\sim \mathcal{N}(\mu + 1.5,1).
\end{equation}

In this scenario:

\begin{equation}\label{eq:auto-091}
n = 100,\mu = 0.50.
\end{equation}

Finally, the scenario \(S10\)considers a negative effect:

\begin{equation}\label{eq:auto-092}
Y_{i}\sim \mathcal{N}( - 0.50,1),n = 50.
\end{equation}

The following table summarizes the simulation plan used.

\textbf{Table 1.} Simulation plan for the one-sample Student\textquotesingle s t-test (B-FBST)

\begingroup\small
\begin{longtable}[]{@{}
  >{\raggedright\arraybackslash}p{(\columnwidth - 8\tabcolsep) * \real{0.1585}}
  >{\raggedright\arraybackslash}p{(\columnwidth - 8\tabcolsep) * \real{0.3165}}
  >{\raggedright\arraybackslash}p{(\columnwidth - 8\tabcolsep) * \real{0.0823}}
  >{\raggedright\arraybackslash}p{(\columnwidth - 8\tabcolsep) * \real{0.1023}}
  >{\raggedright\arraybackslash}p{(\columnwidth - 8\tabcolsep) * \real{0.3403}}@{}}
\toprule\noalign{}
\begin{minipage}[b]{\linewidth}\raggedright
Scenario
\end{minipage} & \begin{minipage}[b]{\linewidth}\raggedright
Distribution
\end{minipage} & \begin{minipage}[b]{\linewidth}\raggedright
\(\mathbf{n}\)
\end{minipage} & \begin{minipage}[b]{\linewidth}\raggedright
\(\mathbf{\mu}\)
\end{minipage} & \begin{minipage}[b]{\linewidth}\raggedright
Objective
\end{minipage} \\
\midrule\noalign{}
\endhead
\bottomrule\noalign{}
\endlastfoot
\(\mathbf{S}\mathbf{1}\) & Normal & 30 & 0.50 & Reference case \\
\(\mathbf{S}\mathbf{2}\) & Student\(t_{3}\) & 30 & 0.50 & Thick tails \\
\(\mathbf{S}\mathbf{3}\) & Contamination & 30 & 0.50 & Extreme values \\
\(\mathbf{S}\mathbf{4}\) & centered log-normal & 50 & 0.50 & Asymmetry \\
\(\mathbf{S}\mathbf{5}\) & Centered nonlinear & 50 & 0.00 & No average effect \\
\(\mathbf{S}\mathbf{6}\) & Large \(n\), small effect & 200 & 0.10 & Weak effect \\
\(\mathbf{S}\mathbf{7}\) & Small sample & 10 & 0.50 & Robustness in small\(n\) \\
\(\mathbf{S}\mathbf{8}\) & Heteroscedasticity & 50 & 0.50 & Non-constant variance \\
\(\mathbf{S}\mathbf{9}\) & Mix of groups & 100 & 0.50 & Heterogeneity \\
\(\mathbf{S}\mathbf{10}\) & Negative effect & 50 & -0.50 & Opposite effect \\
\end{longtable}
\endgroup

This simulation plan allows us to study the behavior of the B-FBST test in contexts favorable and unfavorable to the classical Student\textquotesingle s t-test. It covers, in particular, normality, non-normality, the presence of extreme values, skewness, heteroscedasticity, small samples, and weak or negative effects.

\section{Results of the Student\textquotesingle s t-FBST simulation}

This section presents the results of the Monte Carlo simulation performed to compare three methods applied to the Student\textquotesingle s t-test on a sample: the B-FBST Student\textquotesingle s t-test, the classical Student\textquotesingle s t-test, and the bootstrap Student\textquotesingle s t-test. The results are expressed as the percentage of correct decisions across \(MC = 100\)simulations, with \(R = 200\)bootstrap replications. For the B-FBST Student\textquotesingle s t-test, the decision rule used is:

\begin{equation}\label{eq:auto-093}
EV(H_{0}) < 0.20 \Longrightarrow \text{rejet~de~}H_{0}.
\end{equation}

\begingroup\scriptsize
\begin{longtable}[]{@{}
  >{\raggedright\arraybackslash}p{(\columnwidth - 16\tabcolsep) * \real{0.1158}}
  >{\raggedright\arraybackslash}p{(\columnwidth - 16\tabcolsep) * \real{0.2062}}
  >{\raggedright\arraybackslash}p{(\columnwidth - 16\tabcolsep) * \real{0.0599}}
  >{\raggedright\arraybackslash}p{(\columnwidth - 16\tabcolsep) * \real{0.0466}}
  >{\raggedright\arraybackslash}p{(\columnwidth - 16\tabcolsep) * \real{0.1131}}
  >{\raggedright\arraybackslash}p{(\columnwidth - 16\tabcolsep) * \real{0.0626}}
  >{\raggedright\arraybackslash}p{(\columnwidth - 16\tabcolsep) * \real{0.1255}}
  >{\raggedright\arraybackslash}p{(\columnwidth - 16\tabcolsep) * \real{0.1163}}
  >{\raggedright\arraybackslash}p{(\columnwidth - 16\tabcolsep) * \real{0.1540}}@{}}
\toprule\noalign{}
\begin{minipage}[b]{\linewidth}\raggedright
Scenario
\end{minipage} & \begin{minipage}[b]{\linewidth}\raggedright
Distribution
\end{minipage} & \begin{minipage}[b]{\linewidth}\raggedright
\(\mathbf{n}\)
\end{minipage} & \begin{minipage}[b]{\linewidth}\raggedright
\(\mathbf{\mu}\)
\end{minipage} & \begin{minipage}[b]{\linewidth}\raggedright
B-FBST Student
\end{minipage} & \begin{minipage}[b]{\linewidth}\raggedright
t -test
\end{minipage} & \begin{minipage}[b]{\linewidth}\raggedright
t -test bootstrap
\end{minipage} & \begin{minipage}[b]{\linewidth}\raggedright
average EV
\end{minipage} & \begin{minipage}[b]{\linewidth}\raggedright
p -value average
\end{minipage} \\
\midrule\noalign{}
\endhead
\bottomrule\noalign{}
\endlastfoot
S1 & Normal & 30 & 0.5 & 84 & 79 & 81 & 0.1106 & 0.0472 \\
S2 & Student & 30 & 0.5 & 54 & 44 & 54 & 0.2993 & 0.2068 \\
S3 & Contamination & 30 & 0.5 & 91 & 84 & 92 & 0.0641 & 0.0319 \\
S4 & Log-normal & 50 & 0.5 & 52 & 35 & 53 & 0.2421 & 0.1513 \\
S5 & Non-linear & 50 & 0.0 & 90 & 92 & 90 & 0.5821 & 0.4674 \\
S6 & Large \(n\), small effect & 200 & 0.1 & 38 & 34 & 34 & 0.3725 & 0.2593 \\
S7 & Small sample & 10 & 0.5 & 31 & 25 & 38 & 0.3953 & 0.2678 \\
S8 & Heteroscedasticity & 50 & 0.5 & 72 & 66 & 69 & 0.1599 & 0.0825 \\
S9 & Mix of groups & 100 & 0.5 & 100 & 100 & 100 & 0.000006 & 0.000000008 \\
S10 & Negative effect & 50 & -0.5 & 98 & 96 & 96 & 0.0374 & 0.0098 \\
\end{longtable}
\endgroup

\textbf{Table 2.} Comparative results by scenario

The results in Table (2) and Figure (1) show that the B-FBST Student method exhibits generally favorable performance. It achieves an average of {[}number{]} \(71.0\%\)correct decisions, compared to {[}number \(65.5\%\){]} for the classical Student\textquotesingle s t-test and \(70.7\%\)the bootstrap Student\textquotesingle s t-test. Thus, the proposed method outperforms the classical test on average and remains very close to the bootstrap test.

\begin{figure}[htbp]
\centering
\includegraphics[width=0.95\textwidth,keepaspectratio]{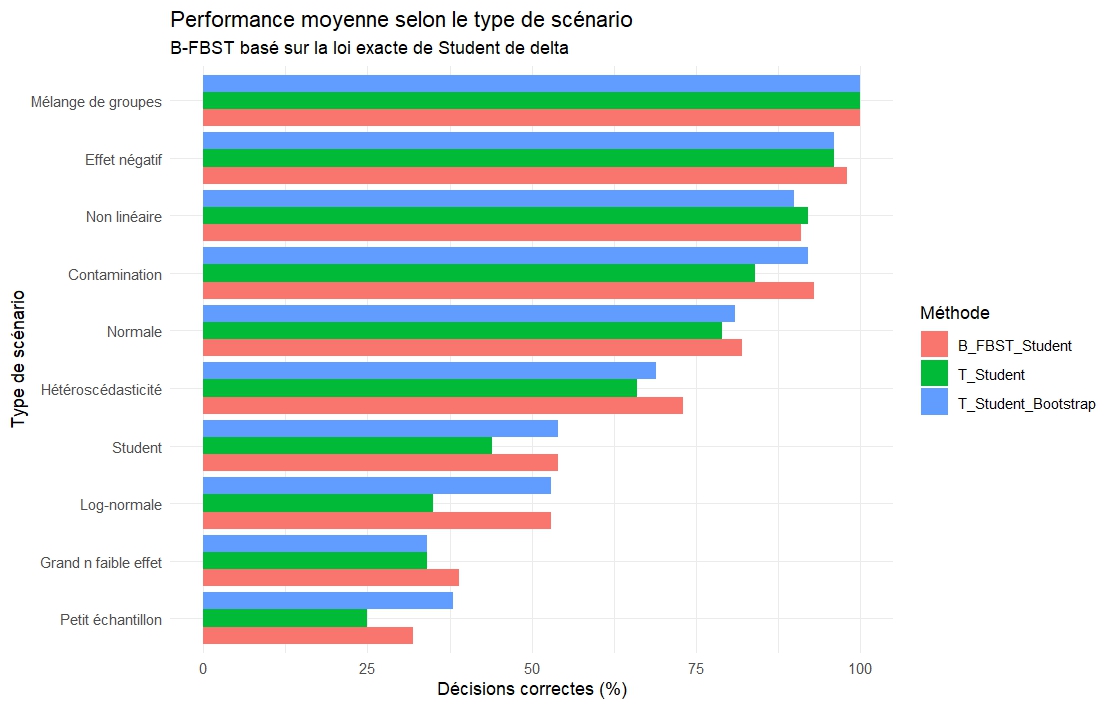}
\caption*{Figure 1. Comparative performance of the B-FBST Student, classical t-test and bootstrap t-test methods according to simulated scenarios}
\end{figure}

\clearpage
\noindent\textbf{Table 3.} Overall performance summary.\par\nopagebreak[4]

\begingroup\small
\begin{longtable}[]{@{}
  >{\raggedright\arraybackslash}p{(\columnwidth - 6\tabcolsep) * \real{0.3485}}
  >{\raggedright\arraybackslash}p{(\columnwidth - 6\tabcolsep) * \real{0.1916}}
  >{\raggedright\arraybackslash}p{(\columnwidth - 6\tabcolsep) * \real{0.2263}}
  >{\raggedright\arraybackslash}p{(\columnwidth - 6\tabcolsep) * \real{0.2336}}@{}}
\toprule\noalign{}
\begin{minipage}[b]{\linewidth}\raggedright
Method
\end{minipage} & \begin{minipage}[b]{\linewidth}\raggedright
Average
\end{minipage} & \begin{minipage}[b]{\linewidth}\raggedright
Minimum
\end{minipage} & \begin{minipage}[b]{\linewidth}\raggedright
Maximum
\end{minipage} \\
\midrule\noalign{}
\endhead
\bottomrule\noalign{}
\endlastfoot
B-FBST Student & 71.0 & 31 & 100 \\
classic t -test & 65.5 & 25 & 100 \\
t -test bootstrap & 70.7 & 34 & 100 \\
\end{longtable}
\endgroup

The summary table confirms that the B-FBST Student\textquotesingle s t-test has the best overall average performance. The bootstrap Student\textquotesingle s t-test also yields good results, with an average very close to that of the B-FBST Student\textquotesingle s t-test. In contrast, the classical Student\textquotesingle s t-test has the lowest average performance, demonstrating its sensitivity in certain non-ideal scenarios.

\textbf{Table 4.} Number of scenarios where each method achieves the best performance

\begingroup\small
\begin{longtable}[]{@{}
  >{\raggedright\arraybackslash}p{(\columnwidth - 2\tabcolsep) * \real{0.4507}}
  >{\raggedright\arraybackslash}p{(\columnwidth - 2\tabcolsep) * \real{0.5493}}@{}}
\toprule\noalign{}
\begin{minipage}[b]{\linewidth}\raggedright
Method
\end{minipage} & \begin{minipage}[b]{\linewidth}\raggedright
Number of scenarios
\end{minipage} \\
\midrule\noalign{}
\endhead
\bottomrule\noalign{}
\endlastfoot
B-FBST Student & 6 \\
classic T-test & 2 \\
T-test bootstrap & 5 \\
\end{longtable}
\endgroup

The B-FBST Student method achieves the best performance in six scenarios. The bootstrap Student test is the best in five scenarios, while the classical test is the best in two scenarios. The total exceeds ten because some methods are tied in certain scenarios, notably in \(S2\), \(S5\)and \(S9\).

Scenario analysis shows that the B-FBST Student test performs particularly well in the normal reference case \(S1\), with \(84\%\)correct decisions compared to those obtained \(79\%\)for the classical test and \(81\%\)the bootstrap test. It also performs well in the thick-tailed scenario \(S2\), where it achieves {[}missing value{]} \(54\%\), compared to only {[}missing value{]} \(44\%\)for the classical test. In the heteroscedasticity scenario \(S8\), it achieves {[}missing value \(72\%\){]}, which is superior to both competing methods. For the negative effect \(S10\), it achieves {[} missing value \(98\%\){]}, compared to those obtained \(96\%\)for the other two methods.

However, the bootstrap test remains slightly superior in some cases. In the contamination scenario \(S3\), it achieves {[}value missing{]} \(92\%\), compared to {[}value missing \(91\%\){]} for the B-FBST Student\textquotesingle s t-test. It is also better in the log-normal scenario \(S4\)and in the small sample size \(S7\). In the nonlinear scenario \(S5\), all three methods give high results, but the classical test achieves the best performance with {[}value missing{]} \(92\%\).

The average e-values are consistent with the logic of the FBST. When the effect is strong, as in \(S9\)and \(S10\), the e-values \(EV(H_{0})\)are very low, indicating strong evidence against the null hypothesis. Conversely, in scenarios \(S5\), , \(S6\)and \(S7\), the average e-values are higher, reflecting weaker or more unstable evidence against \(H_{0}\).

In conclusion, the B-FBST Student\textquotesingle s t-test appears to be a competitive one-sample method for the Student\textquotesingle s t-test. It generally improves upon the classical test and demonstrates robustness in several less-than-ideal situations, particularly in the presence of thick tails, heteroscedasticity, weak effects, or negative effects. However, the bootstrap Student\textquotesingle s t-test sometimes remains slightly superior, especially in cases of contamination, skewness, or small sample sizes.

\section{Conclusion}

This work has led to the proposal of an exact Bayesian formulation of the B-FBST test applied to the Student\textquotesingle s t-test on a single sample. After reviewing the fundamentals of the classical Student\textquotesingle s t-test, the study developed a Bayesian approach based on Jeffreys\textquotesingle{} uninformative prior distribution, resulting in an exact marginal Student\textquotesingle s t-distribution for the standardized parameter under study.

The application of FBST relies on the calculation of the e-value, used as a measure of evidence against the null hypothesis. To improve the stability of the method, bootstrap samples were generated and combined using random weights, resulting in an aggregated posterior density.

The results of the Monte Carlo simulation, performed on 100 replications with 200 bootstrap replications, show that the B-FBST Student\textquotesingle s t-test exhibits generally favorable performance. It achieves an average of 71.0\% correct decisions, compared to 65.5\% for the classical Student\textquotesingle s t-test and 70.7\% for the bootstrap Student\textquotesingle s t-test. These results indicate that the proposed method improves upon the classical test and yields performance very close to that of the bootstrap test.

Scenario analysis also confirms the value of the B-FBST Student\textquotesingle s t-test. In the normal reference case, it achieves 84\% correct decisions, compared to 79\% for the classical test and 81\% for the bootstrap test. In the contaminated data case, it reaches 91\%, compared to 84\% for the classical test, demonstrating its ability to handle extreme values. In the negative effect scenario, it achieves 98\%, compared to 96\% for the other two methods. Finally, in the group mixing case, all three methods achieve 100\%, indicating good detection capabilities when the effect is strong.

Overall, the B-FBST Student\textquotesingle s t-test performed best in 6 scenarios, compared to 5 scenarios for the bootstrap test and 2 scenarios for the classical test, with some situations showing equal results between methods. These results demonstrate that the proposed Bayesian approach is competitive, especially in contexts where the assumptions of the classical Student\textquotesingle s t-test are weak, such as heavy-tailed distributions, heteroscedasticity, weak effects, or the presence of outliers.

However, the results also show that the B-FBST Student\textquotesingle s t-test is not always superior to all other methods. In certain specific cases, such as very small samples, the bootstrap test can yield slightly better results. This does not diminish the value of the B-FBST Student\textquotesingle s t-test, but rather demonstrates that it should be considered a robust and relevant alternative to the classic Student\textquotesingle s t-test, particularly when seeking a method capable of providing a Bayesian interpretation of the statistical evidence.

In conclusion, this study shows that the B-FBST Student test enriches the statistical analysis of the Student\textquotesingle s t-test by providing a Bayesian measure of evidence against the null hypothesis. The results obtained open the way to possible extensions, particularly towards two-sample tests, paired observations, and the study of e-value thresholds better suited to different application contexts.

\section*{Appendices}

\begin{itemize}
\item
  Efron, B. (1979). Bootstrap methods : Another look at the jackknife . \emph{The Annals of Statistics, 7} (1), 1--26. DOI: 10.1214/aos/1176344552.
\item
  Jeffreys, H. (1961). \emph{Theory of probability} (3rd ed .). Oxford University Press .
\item
  Pereira, CADB, \& Stern, J.M. (1999). Evidence and credibility : full Bayesian significance test for precision hypotheses . \emph{Entropy} , \emph{1} (4), 99-110.
\item
  Pereira, CADB, \& Stern, J.M. (2022). The e- value: has fully Bayesian significance measure for precise statistical hypotheses and its research program. \emph{São Paulo Journal of Mathematical Sciences} , \emph{16} (1), 566-584.
\item
  Rubin, D.B. (1981). The Bayesian bootstrap. \emph{The annals of statistics} , 130-134.
\item
  Stern, JM, Pereira, CADB, Lauretto , MDS, Esteves, LG, Izbicki, R., Stern, RB, ... \& Borges, WDS (2025). The e-value and the full Bayesian significance test: logical properties and philosophical consequences . In \emph{Walter Carnielli on Reasoning , Paraconsistency , and Probability} (pp. 487-524). Cham: Springer Nature Switzerland .
\item
  Student. (1908). The probable error of a mean . \emph{Biometrika , 6} (1), 1--25. DOI: 10.1093/ biomet /6.1.1.
\end{itemize}

\section*{Appendices}

\subsection*{Demonstration of the a posteriori law of \(\sigma^{2}\mid y\)}

We consider the normal model with a single sample:

\begin{equation}\label{eq:auto-094}
Y_{1},\ldots,Y_{n} \mid \mu,\sigma^{2}\sim \mathcal{N}(\mu,\sigma^{2}).
\end{equation}

The likelihood of the observed sample \(y = (y_{1},\ldots,y_{n})\)is given by:

\begin{equation}\label{eq:auto-095}
L(\mu,\sigma^{2} \mid y) = (2\pi\sigma^{2})^{- n/2}\exp\left\lbrack - \frac{1}{2\sigma^{2}}\sum_{i = 1}^{n}(y_{i} - \mu)^{2} \right\rbrack.
\end{equation}

We use Jeffreys\textquotesingle{} a priori non-informative law:

\begin{equation}\label{eq:auto-096}
\pi(\mu,\sigma^{2}) \propto \frac{1}{\sigma^{2}}.
\end{equation}

Thus, the joint a posteriori law of \(\mu\)and \(\sigma^{2}\)is proportional to:

\begin{equation}\label{eq:auto-097}
\pi(\mu,\sigma^{2} \mid y) \propto L(\mu,\sigma^{2} \mid y)\pi(\mu,\sigma^{2}).
\end{equation}

SO :

\begin{equation}\label{eq:auto-098}
\pi(\mu,\sigma^{2} \mid y) \propto (\sigma^{2})^{- n/2}\exp\left\lbrack - \frac{1}{2\sigma^{2}}\sum_{i = 1}^{n}(y_{i} - \mu)^{2} \right\rbrack(\sigma^{2})^{- 1}.
\end{equation}

We obtain:

\begin{equation}\label{eq:auto-099}
\pi(\mu,\sigma^{2} \mid y) \propto (\sigma^{2})^{- \frac{n}{2} - 1}\exp\left\lbrack - \frac{1}{2\sigma^{2}}\sum_{i = 1}^{n}(y_{i} - \mu)^{2} \right\rbrack.
\end{equation}

We now use the classical decomposition:

\begin{equation}\label{eq:auto-100}
\sum_{i = 1}^{n}(y_{i} - \mu)^{2} = \sum_{i = 1}^{n}(y_{i} - \bar{y})^{2} + n(\mu - \bar{y})^{2}.
\end{equation}

By asking:

\begin{equation}\label{eq:auto-101}
S = \sum_{i = 1}^{n}(y_{i} - \bar{y})^{2},
\end{equation}

we have :

\begin{equation}\label{eq:auto-102}
\sum_{i = 1}^{n}(y_{i} - \mu)^{2} = S + n(\mu - \bar{y})^{2}.
\end{equation}

So :

\begin{equation}\label{eq:auto-103}
\pi(\mu,\sigma^{2} \mid y) \propto (\sigma^{2})^{- \frac{n}{2} - 1}\exp\left\lbrack - \frac{S + n(\mu - \bar{y})^{2}}{2\sigma^{2}} \right\rbrack.
\end{equation}

We can separate the two terms in the exponential function:

\begin{equation}\label{eq:auto-104}
\pi(\mu,\sigma^{2} \mid y) \propto (\sigma^{2})^{- \frac{n}{2} - 1}\exp\left\lbrack - \frac{S}{2\sigma^{2}} \right\rbrack\exp\left\lbrack - \frac{n(\mu - \bar{y})^{2}}{2\sigma^{2}} \right\rbrack.
\end{equation}

To obtain the marginal distribution of \(\sigma^{2} \mid y\), we integrate with respect to \(\mu\):

\begin{equation}\label{eq:auto-105}
\pi(\sigma^{2} \mid y) \propto (\sigma^{2})^{- \frac{n}{2} - 1}\exp{\left\lbrack - \frac{S}{2\sigma^{2}} \right\rbrack\int_{- \infty}^{+ \infty}{\exp\left\lbrack - \frac{n(\mu - \bar{y})^{2}}{2\sigma^{2}} \right\rbrack}}*d\mu.
\end{equation}

The Gaussian integral is equal to:

\begin{equation}\label{eq:auto-106}
\int_{- \infty}^{+ \infty}{\exp}\left\lbrack - \frac{n(\mu - \bar{y})^{2}}{2\sigma^{2}} \right\rbrack d\mu = \sqrt{\frac{2\pi\sigma^{2}}{n}}.
\end{equation}

This term is proportional to:

\begin{equation}\label{eq:auto-107}
\left( \sigma^{2})^{1/2}. \right.\
\end{equation}

SO :

\begin{equation}\label{eq:auto-108}
\pi(\sigma^{2} \mid y) \propto (\sigma^{2})^{- \frac{n}{2} - 1}(\sigma^{2})^{\frac{1}{2}}\exp\left\lbrack - \frac{S}{2\sigma^{2}} \right\rbrack.
\end{equation}

By simplifying the powers of \(\sigma^{2}\), we obtain:

\begin{equation}\label{eq:auto-109}
- \frac{n}{2} - 1 + \frac{1}{2} = - \frac{n + 1}{2}.
\end{equation}

So :

\begin{equation}\label{eq:auto-110}
\pi(\sigma^{2} \mid y) \propto (\sigma^{2})^{- \frac{n + 1}{2}}\exp\left\lbrack - \frac{S}{2\sigma^{2}} \right\rbrack.,\sigma^{2} > 0.
\end{equation}

Now, the density of an inverse-gamma distribution with parameters \(\alpha\)and \(\beta\)is:

\begin{equation}\label{eq:auto-111}
f(x) = \frac{\beta^{\alpha}}{\Gamma(\alpha)}x^{- (\alpha + 1)}\exp\left( - \frac{\beta}{x} \right),x > 0.
\end{equation}

Compared to:

\begin{equation}\label{eq:auto-112}
\pi(\sigma^{2} \mid y) \propto (\sigma^{2})^{- \frac{n + 1}{2}}\exp\left\lbrack - \frac{S}{2\sigma^{2}} \right\rbrack,
\end{equation}

we identify:

\begin{equation}\label{eq:auto-113}
\alpha + 1 = \frac{n + 1}{2},
\end{equation}

SO :

\begin{equation}\label{eq:auto-114}
\alpha = \frac{n - 1}{2},
\end{equation}

And :

\begin{equation}\label{eq:auto-115}
\beta = \frac{S}{2}.
\end{equation}

Therefore :

\begin{equation}\label{eq:auto-116}
\sigma^{2} \mid y \sim \text{Inverse-Gamma}\left( \frac{n - 1}{2},\frac{S}{2} \right)
\end{equation}

with :

\begin{equation}\label{eq:auto-117}
S = \sum_{i = 1}^{n}(y_{i} - \bar{y})^{2}.
\end{equation}

The exact density is therefore:

\begin{equation}\label{eq:auto-118}
\pi(\sigma^{2} \mid y) = \frac{\left( \frac{S}{2} \right)^{\frac{n - 1}{2}}}{\Gamma\left( \frac{n - 1}{2} \right)}(\sigma^{2})^{- \frac{n + 1}{2}}\exp\left( - \frac{S}{2\sigma^{2}} \right),\sigma^{2} > 0.
\end{equation}

Thus, under Jeffreys\textquotesingle{} prior law \(\pi(\mu,\sigma^{2}) \propto 1/\sigma^{2}\), the posterior marginal law of the variance \(\sigma^{2}\)is an inverse-gamma law with parameters \((n - 1)/2\)and \(S/2\).

\subsection*{Demonstration of the marginal law of \(\mu\mid y\)}

We consider the normal model with a single sample:

\begin{equation}\label{eq:auto-119}
Y_{1},\ldots,Y_{n} \mid \mu,\sigma^{2}\sim \mathcal{N}(\mu,\sigma^{2}).
\end{equation}

Note:

\begin{equation}\label{eq:auto-120}
\bar{y} = \frac{1}{n}\sum_{i = 1}^{n}y_{i}
\end{equation}

And :

\begin{equation}\label{eq:auto-121}
S = \sum_{i = 1}^{n}(y_{i} - \bar{y})^{2}.
\end{equation}

We use Jeffreys\textquotesingle{} prior law:

\begin{equation}\label{eq:auto-122}
\pi(\mu,\sigma^{2}) \propto \frac{1}{\sigma^{2}}.
\end{equation}

The likelihood of the sample is:

\begin{equation}\label{eq:auto-123}
L(\mu,\sigma^{2} \mid y) = (2\pi\sigma^{2})^{- n/2}\exp\left\lbrack - \frac{1}{2\sigma^{2}}\sum_{i = 1}^{n}(y_{i} - \mu)^{2} \right\rbrack.
\end{equation}

Multiplying by the prior distribution, the resulting posterior distribution is proportional to:

\begin{equation}\label{eq:auto-124}
\pi(\mu,\sigma^{2} \mid y) \propto (\sigma^{2})^{- \frac{n}{2} - 1}\exp\left\lbrack - \frac{1}{2\sigma^{2}}\sum_{i = 1}^{n}(y_{i} - \mu)^{2} \right\rbrack.
\end{equation}

We then use the following decomposition:

\begin{equation}\label{eq:auto-125}
\sum_{i = 1}^{n}(y_{i} - \mu)^{2} = \sum_{i = 1}^{n}(y_{i} - \bar{y})^{2} + n(\mu - \bar{y})^{2}.
\end{equation}

SO :

\begin{equation}\label{eq:auto-126}
\sum_{i = 1}^{n}(y_{i} - \mu)^{2} = S + n(\mu - \bar{y})^{2}.
\end{equation}

So :

\begin{equation}\label{eq:auto-127}
\pi(\mu,\sigma^{2} \mid y) \propto (\sigma^{2})^{- \frac{n}{2} - 1}\exp\left\lbrack - \frac{S + n(\mu - \bar{y})^{2}}{2\sigma^{2}} \right\rbrack.
\end{equation}

To obtain the marginal distribution of \(\mu \mid y\), we integrate with respect to \(\sigma^{2}\):

\begin{equation}\label{eq:auto-128}
\pi(\mu \mid y) \propto \int_{0}^{+ \infty}(\sigma^{2})^{- \frac{n}{2} - 1}\exp\left\lbrack - \frac{S + n(\mu - \bar{y})^{2}}{2\sigma^{2}} \right\rbrack d\sigma^{2}.
\end{equation}

We ask:

\begin{equation}\label{eq:auto-129}
A(\mu) = S + n(\mu - \bar{y})^{2}.
\end{equation}

SO :

\begin{equation}\label{eq:auto-130}
\pi(\mu \mid y) \propto \int_{0}^{+ \infty}(\sigma^{2})^{- \frac{n}{2} - 1}\exp\left\lbrack - \frac{A(\mu)}{2\sigma^{2}} \right\rbrack d\sigma^{2}.
\end{equation}

This integral is an inverse-gamma integral. It yields:

\begin{equation}\label{eq:auto-131}
\pi(\mu \mid y) \propto \left\lbrack A(\mu) \right\rbrack^{- \frac{n}{2}}.
\end{equation}

SO :

\begin{equation}\label{eq:auto-132}
\pi(\mu \mid y) \propto \left\lbrack S + n(\mu - \bar{y})^{2} \right\rbrack^{- \frac{n}{2}}.
\end{equation}

As :

\begin{equation}\label{eq:auto-133}
s^{2} = \frac{S}{n - 1},
\end{equation}

we have :

\begin{equation}\label{eq:auto-134}
S = (n - 1)s^{2}.
\end{equation}

So :

\begin{equation}\label{eq:auto-135}
S + n(\mu - \bar{y})^{2} = (n - 1)s^{2}\left\lbrack 1+\frac{n(\mu - \bar{y})^{2}}{(n - 1)s^{2}} \right\rbrack.
\end{equation}

Gold :

\begin{equation}\label{eq:auto-136}
\frac{n(\mu - \bar{y})^{2}}{s^{2}} = \left( \frac{\mu - \bar{y}}{s/\sqrt{n}} \right)^{2}.
\end{equation}

SO :

\begin{equation}\label{eq:auto-137}
\pi(\mu \mid y) \propto \left\lbrack 1+\frac{1}{n - 1}\left( \frac{\mu - \bar{y}}{s/\sqrt{n}} \right)^{2} \right\rbrack^{- \frac{n}{2}}.
\end{equation}

This form corresponds to the probability density function of a Student\textquotesingle s t-distribution with \(n - 1\ \)n degrees of freedom, location parameter \(\bar{y}\), and scale parameter \(s/\sqrt{n}\). Therefore:

\begin{equation}\label{eq:auto-138}
\mu \mid y \sim t_{n - 1}\left( \bar{y},\frac{s}{\sqrt{n}} \right)
\end{equation}

The exact density of \(\mu \mid y\)is:

\begin{equation}\label{eq:auto-139}
\pi(\mu \mid y) = \frac{\Gamma\left( \frac{n}{2} \right)}{\Gamma\left( \frac{n - 1}{2} \right)\sqrt{(n - 1)\pi}\left( \frac{s}{\sqrt{n}} \right)}\left\lbrack 1+\frac{1}{n - 1}\left( \frac{\mu - \bar{y}}{s/\sqrt{n}} \right)^{2} \right\rbrack^{- \frac{n}{2}}.
\end{equation}

Thus, under Jeffreys\textquotesingle{} a priori law:

\begin{equation}\label{eq:auto-140}
\pi(\mu,\sigma^{2}) \propto \frac{1}{\sigma^{2}},
\end{equation}

The posterior marginal distribution of the mean \(\mu\)is a Student\textquotesingle s t-distribution:

\begin{equation}\label{eq:auto-141}
\mu \mid y \sim t_{n - 1}\left( \bar{y},\frac{s}{\sqrt{n}} \right).
\end{equation}

\subsection*{Demonstration of the law of \(\delta\mid y\)}

We start from:

\begin{equation}\label{eq:auto-142}
\mu \mid y \sim t_{n - 1}\left( \bar{y},\frac{s}{\sqrt{n}} \right).
\end{equation}

Let\textquotesingle s state:

\begin{equation}\label{eq:auto-143}
\nu = n - 1,m = \bar{y},a = \frac{s}{\sqrt{n}}.
\end{equation}

The density of a Student\textquotesingle s t-distribution of degree n \(\nu\), center n \(m\), scale n \(a\), is:

\begin{equation}\label{eq:auto-144}
\pi(\mu \mid y) = \frac{\Gamma\left( \frac{\nu + 1}{2} \right)}{\Gamma\left( \frac{\nu}{2} \right)\sqrt{\nu\pi}\,a}\left\lbrack 1+\frac{1}{\nu}\left( \frac{\mu - m}{a} \right)^{2} \right\rbrack^{- \frac{\nu + 1}{2}}.
\end{equation}

Since \(\nu = n - 1\), \(m = \bar{y}\), \(a = s/\sqrt{n}\), we have:

\begin{equation}\label{eq:auto-145}
\pi(\mu \mid y) = \frac{\Gamma\left( \frac{n}{2} \right)}{\Gamma\left( \frac{n - 1}{2} \right)\sqrt{(n - 1)\pi}\,\frac{s}{\sqrt{n}}}\left\lbrack 1+\frac{1}{n - 1}\left( \frac{\mu - \bar{y}}{s/\sqrt{n}} \right)^{2} \right\rbrack^{- \frac{n}{2}}.
\end{equation}

Now, to obtain your density, you need to set:

\begin{equation}\label{eq:auto-146}
\delta = \frac{\mu - \mu_{0}}{s}.
\end{equation}

SO :

\begin{equation}\label{eq:auto-147}
\mu = \mu_{0} + s\delta.
\end{equation}

The Jacobian of the change of variable is:

\begin{equation}\label{eq:auto-148}
\mid \frac{d\mu}{d\delta} \mid = s.
\end{equation}

SO :

\begin{equation}\label{eq:auto-149}
\pi(\delta \mid y) = \pi(\mu_{0} + s\delta \mid y)\,s.
\end{equation}

We replace \(\mu\)with \(\mu_{0} + s\delta\)in the density of \(\mu \mid y\):

\begin{equation}\label{eq:auto-150}
\pi\left( \delta\mid y \right) = \frac{\Gamma\left( \frac{n}{2} \right)}{\Gamma\left( \frac{n - 1}{2} \right)\sqrt{(n - 1)\pi}\,\frac{s}{\sqrt{n}}}\left\lbrack 1+\frac{1}{n - 1}\left( \frac{\mu_{0} + s\delta - \bar{y}}{\frac{s}{\sqrt{n}}} \right)^{2} \right\rbrack^{- \frac{n}{2}}s.
\end{equation}

The factor \(s\)simplifies:

\begin{equation}\label{eq:auto-151}
\frac{s}{s/\sqrt{n}} = \sqrt{n} = \frac{1}{1/\sqrt{n}}.
\end{equation}

SO :

\begin{equation}\label{eq:auto-152}
\pi(\delta \mid y) = \frac{\Gamma\left( \frac{n}{2} \right)}{\Gamma\left( \frac{n - 1}{2} \right)\sqrt{(n - 1)\pi}\,\frac{1}{\sqrt{n}}}\left\lbrack 1+\frac{1}{n - 1}\left( \frac{\mu_{0} + s\delta - \bar{y}}{s/\sqrt{n}} \right)^{2} \right\rbrack^{- \frac{n}{2}}.
\end{equation}

Gold :

\begin{equation}\label{eq:auto-153}
\mu_{0} + s\delta - \bar{y} = s\delta - (\bar{y} - \mu_{0}).
\end{equation}SO :

\begin{equation}\label{eq:auto-154}
\frac{\mu_{0} + s\delta - \bar{y}}{s/\sqrt{n}} = \frac{s\delta - (\bar{y} - \mu_{0})}{s/\sqrt{n}}.
\end{equation}

We divide by \(s\):

\begin{equation}\label{eq:auto-155}
\frac{s\delta - (\bar{y} - \mu_{0})}{s/\sqrt{n}} = \frac{\delta - \frac{\bar{y} - \mu_{0}}{s}}{1/\sqrt{n}}.
\end{equation}

We ask:

\begin{equation}\label{eq:auto-156}
\widehat{\delta} = \frac{\bar{y} - \mu_{0}}{s}.
\end{equation}

So :

\begin{equation}\label{eq:auto-157}
\frac{\mu_{0} + s\delta - \bar{y}}{s/\sqrt{n}} = \frac{\delta - \widehat{\delta}}{1/\sqrt{n}}.
\end{equation}

So finally:

\begin{equation}\label{eq:auto-158}
\pi(\delta \mid y) = \frac{\Gamma\left( \frac{n}{2} \right)}{\Gamma\left( \frac{n - 1}{2} \right)\sqrt{(n - 1)\pi}\,\frac{1}{\sqrt{n}}}\left\lbrack 1+\frac{1}{n - 1}\left( \frac{\delta - \widehat{\delta}}{1/\sqrt{n}} \right)^{2} \right\rbrack^{- \frac{n}{2}}
\end{equation}

with :

\begin{equation}\label{eq:auto-159}
\widehat{\delta} = \frac{\bar{y} - \mu_{0}}{s}
\end{equation}

SO :

\begin{equation}\label{eq:auto-160}
\delta \mid y \sim t_{n - 1}\left( \widehat{\delta},\frac{1}{\sqrt{n}} \right)
\end{equation}

\subsection*{Mathematical justification of the construction and aggregation of the B-FBST Student method}

The proposed B-FBST Student method relies on combining a posteriori densities obtained from bootstrap replications, using random weights following a Dirichlet distribution, and then applying the geometric principle of the Full Bayesian Significance Test (FBST) to the aggregate density. Its construction is inspired by the principle of random weighting introduced by Rubin (1981) within the Bayesian Bootstrap framework.

\subsubsection*{a. Bootstrap construction of densities a posteriori}

For each bootstrap sample \(y^{*(b)}\), where \(b = 1,\ldots,R\), we calculate the empirical mean and standard deviation:

\begin{equation}\label{eq:auto-161}
{\overline{y}}^{*(b)} = \frac{1}{n}\sum_{i = 1}^{n}y_{i}^{*(b)}
\end{equation}

\begin{equation}\label{eq:auto-162}
s_{b}^{*} = \sqrt{\frac{1}{n - 1}\sum_{i = 1}^{n}\left( y_{i}^{*(b)} - {\overline{y}}^{*(b)} \right)^{2}}.
\end{equation}

The standardized parameter is defined, for each replication, by:

\begin{equation}\label{eq:auto-163}
\delta^{*(b)} = \frac{\mu - \mu_{0}}{s_{b}^{*}}.
\end{equation}

Its estimator is:

\begin{equation}\label{eq:auto-164}
{\widehat{\delta}}_{b}^{*} = \frac{{\overline{y}}^{*(b)} - \mu_{0}}{s_{b}^{*}}.
\end{equation}

Under the normal model and the chosen Jeffreys prior law, the distribution used for each bootstrap replication is a Student\textquotesingle s t-distribution with n \(n - 1\)degrees of freedom:

\begin{equation}\label{eq:auto-165}
\delta^{*(b)} \mid y^{*(b)} \sim t_{n - 1}\left( {\widehat{\delta}}_{b}^{*},\frac{1}{\sqrt{n}} \right).
\end{equation}

Its density is written as:

Each density thus obtained is positive and normalized:

\begin{equation}\label{eq:auto-166}
\pi^{*(b)}\left( \delta \mid y^{*(b)} \right) \geq 0
\end{equation}

And :

\begin{equation}\label{eq:auto-167}
\int_{- \infty}^{+ \infty}\pi^{*(b)}\left( \delta \mid y^{*(b)} \right)\, d\delta = 1.
\end{equation}

Each replication thus provides its own probability density.

\subsubsection*{b. Aggregation of densities using Dirichlet weighting}

The weighting of bootstrap contributions is inspired by the principle of Bayesian Bootstrap introduced by Rubin (1981), in which the uncertainty associated with the empirical distribution is represented by random weights following a Dirichlet law.

In the present method, this principle is transposed to the level of a posteriori densities resulting from bootstrap replications. The weights are defined by:

\begin{equation}\label{eq:auto-168}
\left( W_{1},\ldots,W_{R} \right) \sim {Dirichlet}(1,\ldots,1),
\end{equation}

with :

\begin{equation}\label{eq:auto-169}
W_{b} \geq 0,\quad\quad\sum_{b = 1}^{R}W_{b} = 1.
\end{equation}

The aggregate density of the B-FBST method is then defined by:

\begin{equation}\label{eq:auto-170}
{\widetilde{\pi}}_{B}(\delta \mid y) = \sum_{b = 1}^{R}W_{b}\pi^{*(b)}\left( \delta \mid y^{*(b)} \right)
\end{equation}

The validity of this construction relies on the positivity of the weights and their unit sum. Indeed, using the linearity of integration, we obtain:

\begin{equation}\label{eq:auto-171}
\begin{aligned}
\int_{- \infty}^{+ \infty}{\widetilde{\pi}}_{B}(\delta \mid y)\, d\delta & = \int_{- \infty}^{+ \infty}{\sum_{b = 1}^{R}W_{b}}\pi^{*(b)}\left( \delta \mid y^{*(b)} \right)\, d\delta \\
 & = \sum_{b = 1}^{R}W_{b}\int_{- \infty}^{+ \infty}\pi^{*(b)}\left( \delta \mid y^{*(b)} \right)\, d\delta \\
 & = \sum_{b = 1}^{R}W_{b} \\
 & = 1.
\end{aligned}
\end{equation}

Therefore :

\begin{equation}\label{eq:auto-172}
\int_{- \infty}^{+ \infty}{\widetilde{\pi}}_{B}(\delta \mid y)\, d\delta = 1
\end{equation}

The aggregate density is therefore a proper, positive and normalized probability density.

Furthermore, the properties of Dirichlet\textquotesingle s law allow us to establish that:

\begin{equation}\label{eq:auto-173}
E\left( W_{b} \right) = \frac{1}{R}
\end{equation}

And :

\begin{equation}\label{eq:auto-174}
{Var}\left( W_{b} \right) = \frac{R - 1}{R^{2}(R + 1)}.
\end{equation}

For two distinct indices \(b \neq j\), the covariance is:

\begin{equation}\label{eq:auto-175}
{Cov}\left( W_{b},W_{j} \right) = - \frac{1}{R^{2}(R + 1)}.
\end{equation}

It follows that the expected value of the aggregate density, conditional on bootstrap replications, is:

\begin{equation}\label{eq:auto-176}
\begin{aligned}
E_{W}\left\lbrack {\widetilde{\pi}}_{B}(\delta \mid y) \right\rbrack & = \sum_{b = 1}^{R}E\left( W_{b} \right)\pi^{*(b)}\left( \delta \mid y^{*(b)} \right) \\
 & = \frac{1}{R}\sum_{b = 1}^{R}\pi^{*(b)}\left( \delta \mid y^{*(b)} \right).
\end{aligned}
\end{equation}

So :

\begin{equation}\label{eq:auto-177}
E_{W}\left\lbrack {\widetilde{\pi}}_{B}(\delta \mid y) \right\rbrack = \frac{1}{R}\sum_{b = 1}^{R}\pi^{*(b)}\left( \delta \mid y^{*(b)} \right)
\end{equation}

This property shows that Dirichlet weighting maintains a uniform mixture in expectation, while introducing additional variability due to the randomization of the weights. More precisely:

\begin{equation}\label{eq:auto-178}
{\ \ \ \ \ \ \ \ \ \ \ \ \ \ \ \ \ \ \ \ \ \ \ \ \ \ \ \ Var}_{W}\left\lbrack {\widetilde{\pi}}_{B}(\delta \mid y) \right\rbrack = \frac{1}{R(R + 1)}\sum_{b = 1}^{R}\left\lbrack \pi^{*(b)}\left( \delta \mid y^{*(b)} \right) - {\overline{\pi}}_{R}(\delta) \right\rbrack^{2}
\end{equation}

Or :

\begin{equation}\label{eq:auto-179}
{\overline{\pi}}_{R}(\delta) = \frac{1}{R}\sum_{b = 1}^{R}\pi^{*(b)}\left( \delta \mid y^{*(b)} \right).
\end{equation}

This expression characterizes the variability of the aggregate density induced by the Dirichlet weights.

\subsubsection*{c. Construction of the tangential region and calculation of the e-value}

The initial null hypothesis:

\begin{equation}\label{eq:auto-180}
H_{0}:\mu = \mu_{0}
\end{equation}

corresponds, after standardization, to:

\begin{equation}\label{eq:auto-181}
H_{0}:\delta = 0.
\end{equation}

By applying the geometric principle of FBST to the aggregate density, the tangential region of the proposed method is defined by:

\begin{equation}\label{eq:auto-182}
T_{B} = \left\{ \delta:{\widetilde{\pi}}_{B}(\delta \mid y) > {\widetilde{\pi}}_{B}(0 \mid y) \right\}.
\end{equation}

The e-value associated with the null hypothesis is then given by:

\begin{equation}\label{eq:auto-183}
EV_{B}\left( H_{0} \right) = 1 - \int_{T_{B}}^{}{\widetilde{\pi}}_{B}(\delta \mid y)\, d\delta
\end{equation}

Since the aggregate density is positive and normalized, this quantity necessarily satisfies:

\begin{equation}\label{eq:auto-184}
0 \leq EV_{B}\left( H_{0} \right) \leq 1
\end{equation}

The B-FBST procedure thus provides a measure of evidence constructed from the tangential region associated with the aggregate density. It adopts the geometric principle of FBST while integrating the contributions of bootstrap replications according to a random Dirichlet weighting.

This construction represents a hybrid extension of the FBST principle. The resulting mixture is used as the aggregate density in the procedure, without being considered the exact posterior distribution of the initial model. The statistical properties of this new procedure, particularly its calibration under the null hypothesis, its power, and its stability, are examined in the Monte Carlo simulation study.

\subsection*{Convergence in probability of the B-FBST aggregate density}

The objective of this demonstration is to establish that the aggregate density of the B-FBST method converges, as the number of bootstrap replications \(R\)increases, towards the average density induced by the bootstrap mechanism. In other words, we seek to demonstrate that:

\begin{equation}\label{eq:auto-185}
{\widetilde{\pi}}_{B,R}(\delta)\overset{P}{\underset{R \rightarrow \infty}{\rightarrow}}\pi_{boot}(\delta \mid y),
\end{equation}

where the limiting density is defined by:

\begin{equation}\label{eq:auto-186}
\pi_{boot}(\delta \mid y) = E_{*}\left\lbrack \pi^{*}\left( \delta \mid Y^{*} \right) \mid y \right\rbrack.
\end{equation}

To establish this result, we recall that the aggregate density of the B-FBST method is given by:

\begin{equation}\label{eq:bootstrap-aggregate}
{\widetilde{\pi}}_{B,R}(\delta) = \sum_{b = 1}^{R}W_{b}\pi_{b}^{*}(\delta),
\end{equation}

where the weights follow a Dirichlet law:

\begin{equation}\label{eq:auto-188}
\left( W_{1},\ldots,W_{R} \right) \sim {Dirichlet}(1,\ldots,1),
\end{equation}

with :

\begin{equation}\label{eq:auto-189}
W_{b} \geq 0,\quad\quad\sum_{b = 1}^{R}W_{b} = 1.
\end{equation}

A fundamental property of Dirichlet\textquotesingle s law allows these weights to be represented in the form:

\begin{equation}\label{eq:auto-190}
W_{b} = \frac{G_{b}}{\sum_{j = 1}^{R}G_{j}},
\end{equation}

where the variables \(G_{1},\ldots,G_{R}\)are independent and follow an exponential distribution with parameter 1:

\begin{equation}\label{eq:auto-191}
G_{b}\overset{iid}{\sim}{Exp}(1).
\end{equation}

These variables satisfy:

\begin{equation}\label{eq:auto-192}
E\left( G_{b} \right) = 1,\quad\quad{Var}\left( G_{b} \right) = 1.
\end{equation}

Substituting this representation into expression~\eqref{eq:bootstrap-aggregate}, we obtain:

\begin{equation}\label{eq:auto-193}
{\widetilde{\pi}}_{B,R}(\delta) = \frac{\sum_{b = 1}^{R}G_{b}\pi_{b}^{*}(\delta)}{\sum_{b = 1}^{R}G_{b}}.
\end{equation}

Dividing the numerator and denominator by \(R\), this expression becomes:

\begin{equation}\label{eq:ratio-expression}
{\widetilde{\pi}}_{B,R}(\delta) = \frac{\frac{1}{R}\sum_{b = 1}^{R}G_{b}\pi_{b}^{*}(\delta)}{\frac{1}{R}\sum_{b = 1}^{R}G_{b}}.
\end{equation}

First, convergence of the denominator. Since the variables \(G_{b}\)are independent, identically distributed, and have a finite expected value, the strong law of large numbers gives:

\begin{equation}\label{eq:denominator-limit}
\frac{1}{R}\sum_{b = 1}^{R}G_{b}\overset{P}{\underset{R \rightarrow \infty}{\rightarrow}}E\left( G_{b} \right) = 1.
\end{equation}

Secondly, convergence of the numerator. Consider the variables:

\begin{equation}\label{eq:auto-196}
X_{b}(\delta) = G_{b}\pi_{b}^{*}(\delta).
\end{equation}

Conditional on the initial sample \(y\), the variables \(X_{b}(\delta)\)are independent and identically distributed, since the bootstrap replications and Gamma variables are generated independently. Furthermore, the variables \(G_{b}\)are independent of the bootstrap densities. Thus:

Furthermore, the Student\textquotesingle s t-distribution densities used in the B-FBST method are uniformly bounded for a fixed sample size \(n\). Therefore:

\begin{equation}\label{eq:auto-197}
E\left\lbrack |X_{b}(\delta)| \mid y \right\rbrack < \infty.
\end{equation}

The conditions for applying the strong law of large numbers are therefore satisfied. We obtain:

\begin{equation}\label{eq:numerator-limit}
\frac{1}{R}\sum_{b = 1}^{R}G_{b}\pi_{b}^{*}(\delta)\overset{P}{\underset{R \rightarrow \infty}{\rightarrow}}\pi_{boot}(\delta \mid y).
\end{equation}

Third, convergence of the aggregate density. By combining results~\eqref{eq:denominator-limit} and \eqref{eq:numerator-limit}, and using the convergence theorem for the ratio, we deduce from expression~\eqref{eq:ratio-expression} that:

\begin{equation}\label{eq:auto-199}
\begin{aligned}
{\widetilde{\pi}}_{B,R}(\delta) & = \frac{\frac{1}{R}\sum_{b = 1}^{R}G_{b}\pi_{b}^{*}(\delta)}{\frac{1}{R}\sum_{b = 1}^{R}G_{b}} \\
 & \overset{P}{\underset{R \rightarrow \infty}{\rightarrow}}\frac{\pi_{boot}(\delta \mid y)}{1}.
\end{aligned}
\end{equation}

Finally, we obtain:

\begin{equation}\label{eq:auto-200}
{\widetilde{\pi}}_{B,R}(\delta)\overset{P}{\underset{R \rightarrow \infty}{\rightarrow}}\pi_{boot}(\delta \mid y).
\end{equation}

This demonstration establishes that the aggregate density of the B-FBST method converges in probability to the average density induced by the bootstrap mechanism as the number of replications increases. Thus, the random Dirichlet weighting allows the combination of different bootstrap posterior densities while preserving an asymptotic limit determined by their conditional expectation. This result justifies the asymptotic stability of the proposed construction with respect to the number of bootstrap replications. It should be noted, however, that the limiting density represents an average distribution induced by the bootstrap and does not necessarily correspond to the exact posterior distribution of the initial statistical model.

\subsection*{R Code}

\begin{Verbatim}[fontsize=\footnotesize,breaklines=true,breakanywhere=true]
# ============================================================
# B-FBST Student exact vs t-test classique vs t-test bootstrap
# MC = 100, R = 200
# ============================================================

library(rjags)
library(coda)
library(ggplot2)
library(dplyr)
library(tidyr)

# ============================================================
# 1. Poids de Rubin : Dirichlet(1,...,1)
# ============================================================

dirichlet_weights_jags <- function(R){
  
  model_string <- "
  model{
    w[1:R] ~ ddirch(alpha[])
    for(i in 1:R){
      alpha[i] <- 1
    }
  }
  "
  
  jm <- jags.model(
    textConnection(model_string),
    data = list(R = R),
    n.chains = 1,
    quiet = TRUE
  )
  
  samp <- coda.samples(jm, variable.names = "w", n.iter = 1)
  as.numeric(samp[[1]][1, ])
}

# ============================================================
# 2. Loi exacte : delta | y* ~ t_{n-1}(delta_hat*, 1/sqrt(n))
# ============================================================

posterior_delta_student <- function(delta_hat, n, grid){
  
  nu <- n - 1
  scale_delta <- 1 / sqrt(n)
  
  dens <- dt((grid - delta_hat) / scale_delta, df = nu) / scale_delta
  
  h <- grid[2] - grid[1]
  dens <- dens / sum(dens * h)
  
  return(dens)
}

# ============================================================
# 3. B-FBST Student exact
# ============================================================

b_fbst_student <- function(y, mu0 = 0,
                           R = 200,
                           alpha_ev = 0.20,
                           grid_size = 2000){
  
  n <- length(y)
  grid <- seq(-5, 5, length.out = grid_size)
  h <- grid[2] - grid[1]
  
  post_mat <- matrix(0, nrow = R, ncol = grid_size)
  
  for(b in 1:R){
    
    yb <- sample(y, n, replace = TRUE)
    sb <- sd(yb)
    
    if(is.na(sb) || sb == 0){
      delta_hat_b <- 0
    } else {
      delta_hat_b <- (mean(yb) - mu0) / sb
    }
    
    post_mat[b, ] <- posterior_delta_student(
      delta_hat = delta_hat_b,
      n = n,
      grid = grid
    )
  }
  
  w <- dirichlet_weights_jags(R)
  
  post_rubin <- as.numeric(w %*% post_mat)
  post_rubin <- post_rubin / sum(post_rubin * h)
  
  pi0 <- post_rubin[which.min(abs(grid - 0))]
  
  T_region <- post_rubin > pi0
  
  EV <- 1 - sum(post_rubin[T_region] * h)
  
  reject <- ifelse(EV < alpha_ev, 1, 0)
  
  return(list(
    EV = EV,
    reject = reject
  ))
}

# ============================================================
# 4. t-test classique
# ============================================================

ttest_classique <- function(y, mu0 = 0, alpha = 0.05){
  
  test <- t.test(y, mu = mu0)
  
  reject <- ifelse(test$p.value < alpha, 1, 0)
  
  return(list(
    p_value = test$p.value,
    reject = reject
  ))
}

# ============================================================
# 5. t-test bootstrap classique
# ============================================================

ttest_bootstrap <- function(y, mu0 = 0, R = 200, alpha = 0.05){
  
  n <- length(y)
  mean_star <- numeric(R)
  
  for(b in 1:R){
    yb <- sample(y, n, replace = TRUE)
    mean_star[b] <- mean(yb)
  }
  
  IC <- quantile(
    mean_star,
    probs = c(alpha/2, 1 - alpha/2),
    na.rm = TRUE
  )
  
  reject <- ifelse(IC[1] > mu0 | IC[2] < mu0, 1, 0)
  
  return(list(
    IC_low = IC[1],
    IC_high = IC[2],
    reject = reject
  ))
}

# ============================================================
# 6. Génération des 10 scénarios
# ============================================================

generate_data_t <- function(scenario, n, mu_true = 0){
  
  if(scenario == "S1"){
    y <- rnorm(n, mean = mu_true, sd = 1)
  }
  
  if(scenario == "S2"){
    y <- mu_true + rt(n, df = 3)
  }
  
  if(scenario == "S3"){
    eps <- 0.10
    g <- rbinom(n, 1, eps)
    y <- rnorm(n, mean = mu_true, sd = 1)
    y[g == 1] <- rnorm(sum(g == 1), mean = mu_true + 5, sd = 3)
  }
  
  if(scenario == "S4"){
    z <- rnorm(n)
    y <- exp(z) - exp(0.5) + mu_true
  }
  
  if(scenario == "S5"){
    x <- rnorm(n)
    y <- x^2 - 1 + mu_true + rnorm(n, 0, 0.5)
  }
  
  if(scenario == "S6"){
    y <- rnorm(n, mean = mu_true, sd = 1)
  }
  
  if(scenario == "S7"){
    y <- rnorm(n, mean = mu_true, sd = 1)
  }
  
  if(scenario == "S8"){
    x <- rnorm(n)
    y <- mu_true + sqrt(1 + x^2) * rnorm(n)
  }
  
  if(scenario == "S9"){
    g <- rbinom(n, 1, 0.30)
    y <- numeric(n)
    y[g == 0] <- rnorm(sum(g == 0), mean = mu_true, sd = 1)
    y[g == 1] <- rnorm(sum(g == 1), mean = mu_true + 1.5, sd = 1)
  }
  
  if(scenario == "S10"){
    y <- rnorm(n, mean = mu_true, sd = 1)
  }
  
  return(y)
}

# ============================================================
# 7. Plan des scénarios
# ============================================================

plan <- data.frame(
  scenario = c("S1", "S2", "S3", "S4", "S5",
               "S6", "S7", "S8", "S9", "S10"),
  
  distribution = c("Normale", "Student", "Contamination",
                   "Log-normale", "Non linéaire",
                   "Grand n faible effet", "Petit échantillon",
                   "Hétéroscédasticité", "Mélange de groupes",
                   "Effet négatif"),
  
  n = c(30, 30, 30, 50, 50, 200, 10, 50, 100, 50),
  
  mu_true = c(0.5, 0.5, 0.5, 0.5, 0,
              0.10, 0.5, 0.5, 0.5, -0.5)
)

# ============================================================
# 8. Simulation Monte Carlo
# ============================================================

run_simulation <- function(MC = 100,
                           R = 200,
                           mu0 = 0,
                           alpha_ev = 0.20){
  
  results <- data.frame()
  
  for(s in 1:nrow(plan)){
    
    scenario <- plan$scenario[s]
    distribution <- plan$distribution[s]
    n <- plan$n[s]
    mu_true <- plan$mu_true[s]
    
    correct_bfbst <- numeric(MC)
    correct_ttest <- numeric(MC)
    correct_boot <- numeric(MC)
    
    EV_values <- numeric(MC)
    p_values <- numeric(MC)
    
    for(m in 1:MC){
      
      y <- generate_data_t(scenario, n, mu_true)
      
      bfbst <- b_fbst_student(
        y = y,
        mu0 = mu0,
        R = R,
        alpha_ev = alpha_ev
      )
      
      tclassic <- ttest_classique(
        y = y,
        mu0 = mu0
      )
      
      tboot <- ttest_bootstrap(
        y = y,
        mu0 = mu0,
        R = R
      )
      
      reject_bfbst <- bfbst$reject
      reject_ttest <- tclassic$reject
      reject_boot <- tboot$reject
      
      EV_values[m] <- bfbst$EV
      p_values[m] <- tclassic$p_value
      
      if(mu_true == mu0){
        correct_bfbst[m] <- ifelse(reject_bfbst == 0, 1, 0)
        correct_ttest[m] <- ifelse(reject_ttest == 0, 1, 0)
        correct_boot[m] <- ifelse(reject_boot == 0, 1, 0)
      } else {
        correct_bfbst[m] <- ifelse(reject_bfbst == 1, 1, 0)
        correct_ttest[m] <- ifelse(reject_ttest == 1, 1, 0)
        correct_boot[m] <- ifelse(reject_boot == 1, 1, 0)
      }
    }
    
    results <- rbind(results, data.frame(
      scenario = scenario,
      distribution = distribution,
      n = n,
      mu_true = mu_true,
      B_FBST_Student = mean(correct_bfbst) * 100,
      T_Student = mean(correct_ttest) * 100,
      T_Student_Bootstrap = mean(correct_boot) * 100,
      mean_EV = mean(EV_values),
      mean_pvalue = mean(p_values)
    ))
    
    print(results)
  }
  
  return(results)
}

# ============================================================
# 9. Exécution
# ============================================================

set.seed(123)

res_ttest <- run_simulation(
  MC = 100,
  R = 200,
  mu0 = 0,
  alpha_ev = 0.20
)

print(res_ttest)

# ============================================================
# 10. Tableau résumé
# ============================================================

resume_tableau <- res_ttest %>%
  summarise(
    moyenne_B_FBST = mean(B_FBST_Student),
    moyenne_T_Student = mean(T_Student),
    moyenne_T_Student_Bootstrap = mean(T_Student_Bootstrap),
    
    min_B_FBST = min(B_FBST_Student),
    min_T_Student = min(T_Student),
    min_T_Student_Bootstrap = min(T_Student_Bootstrap),
    
    max_B_FBST = max(B_FBST_Student),
    max_T_Student = max(T_Student),
    max_T_Student_Bootstrap = max(T_Student_Bootstrap)
  )

print(resume_tableau)

# ============================================================
# 11. Nombre de fois où chaque méthode est la meilleure
# ============================================================

nombre_meilleure_methode <- res_ttest %>%
  select(scenario, B_FBST_Student, T_Student, T_Student_Bootstrap) %>%
  pivot_longer(
    cols = c(B_FBST_Student, T_Student, T_Student_Bootstrap),
    names_to = "methode",
    values_to = "score"
  ) %>%
  group_by(scenario) %>%
  filter(score == max(score)) %>%
  ungroup() %>%
  count(methode)

print(nombre_meilleure_methode)

# ============================================================
# 12. Tableau long pour ggplot2
# ============================================================

res_long <- res_ttest %>%
  select(scenario, distribution,
         B_FBST_Student, T_Student, T_Student_Bootstrap) %>%
  pivot_longer(
    cols = c(B_FBST_Student, T_Student, T_Student_Bootstrap),
    names_to = "methode",
    values_to = "decisions_correctes"
  )

# ============================================================
# 13. Graphe 1 : décisions correctes par scénario
# ============================================================

ggplot(res_long, aes(x = scenario,
                     y = decisions_correctes,
                     fill = methode)) +
  geom_col(position = "dodge") +
  labs(
    title = "Comparaison B-FBST Student, t-test classique et t-test bootstrap",
    subtitle = "Nombre moyen de décisions correctes sur 100 simulations",
    x = "Scénario",
    y = "Décisions correctes (%)",
    fill = "Méthode"
  ) +
  theme_minimal()

# ============================================================
# 14. Graphe 2 : performance selon le type de scénario
# ============================================================

ggplot(res_long, aes(x = reorder(distribution, decisions_correctes),
                     y = decisions_correctes,
                     fill = methode)) +
  geom_col(position = "dodge") +
  coord_flip() +
  labs(
    title = "Performance moyenne selon le type de scénario",
    subtitle = "B-FBST basé sur la loi exacte de Student de delta",
    x = "Type de scénario",
    y = "Décisions correctes (%)",
    fill = "Méthode"
  ) +
  theme_minimal()

# ============================================================
# 15. Graphe 3 : gain relatif du B-FBST
# ============================================================

res_diff <- res_ttest %>%
  mutate(
    diff_BF_T = B_FBST_Student - T_Student,
    diff_BF_Boot = B_FBST_Student - T_Student_Bootstrap
  ) %>%
  select(scenario, diff_BF_T, diff_BF_Boot) %>%
  pivot_longer(
    cols = c(diff_BF_T, diff_BF_Boot),
    names_to = "comparaison",
    values_to = "difference"
  )

ggplot(res_diff, aes(x = scenario,
                     y = difference,
                     fill = comparaison)) +
  geom_col(position = "dodge") +
  geom_hline(yintercept = 0, linetype = "dashed") +
  labs(
    title = "Gain relatif du B-FBST Student",
    subtitle = "Différence en décisions correctes (%)",
    x = "Scénario",
    y = "Différence (%)",
    fill = "Comparaison"
  ) +
  theme_minimal()
\end{Verbatim}

\end{document}